\pdfoutput=1
\RequirePackage{ifpdf}
\ifpdf 
\documentclass[pdftex]{sigma}
\else
\documentclass{sigma}
\fi

\begin{document}

\newcommand{\arXivNumber}{1504.00558}

\allowdisplaybreaks

\renewcommand{\thefootnote}{$\star$}

\renewcommand{\PaperNumber}{050}

\FirstPageHeading

\ShortArticleName{Embeddings of the Racah Algebra into the Bannai--Ito Algebra}

\ArticleName{Embeddings of the Racah Algebra\\ into the Bannai--Ito Algebra\footnote{This paper is a~contribution to the Special Issue on Exact Solvability and Symmetry Avatars
in honour of Luc Vinet.
The full collection is available at
\href{http://www.emis.de/journals/SIGMA/ESSA2014.html}{http://www.emis.de/journals/SIGMA/ESSA2014.html}}}

\Author{Vincent X.~GENEST, Luc VINET and Alexei ZHEDANOV}
\AuthorNameForHeading{V.X.~Genest, L.~Vinet and A.~Zhedanov}
\Address{Centre de Recherches Math\'ematiques, Universit\'e de Montr\'eal,\\ C.P. 6128, Succ.~Centre-Ville, Montr\'eal, QC, Canada, H3C 3J7}

\Email{\href{mailto:genestvi@crm.umontreal.ca}{genestvi@crm.umontreal.ca}, \href{mailto:vinetl@crm.umontreal.ca}{vinetl@crm.umontreal.ca}, \href{mailto:zhedanov@yahoo.com}{zhedanov@yahoo.com}}

\ArticleDates{Received April 02, 2015, in f\/inal form June 25, 2015; Published online June 30, 2015}

\Abstract{Embeddings of the Racah algebra into the Bannai--Ito algebra are proposed in two realizations. First, quadratic combinations of the Bannai--Ito algebra generators in their standard realization on the space of polynomials are seen to generate a central extension of the Racah algebra. The result is also seen to hold independently of the realization. \mbox{Second}, the relationship between the realizations of the Bannai--Ito and Racah algebras by the intermediate Casimir operators of the $\mathfrak{osp}(1|2)$ and $\mathfrak{su}(1,1)$ Racah problems is established. Equivalently, this gives an embedding of the invariance algebra of the generic super\-integrable system on the two-sphere into the invariance algebra of its extension with ref\/lections, which are respectively isomorphic to the Racah and Bannai--Ito algebras.}


\Classification{33C80}

\renewcommand{\thefootnote}{\arabic{footnote}}
\setcounter{footnote}{0}

\section{Introduction}
The purpose of this paper is to examine the relationship between two algebraic structures that have received some attention in the literature as of late: the Racah algebra and the Bannai--Ito algebra. Using two realizations of these algebras, it will be shown that the former can be embedded into the latter.

The Racah algebra, in its standard presentation \eqref{Racah-1}, f\/irst appeared in 1988 in the paper \cite{1988_Granovskii&Zhedanov_JETP_94_49} by Granovskii and Zhedanov, who used this algebraic structure to study the symmetry group of the $6j$ symbols that arise from the coupling of three angular momenta, also known as the \emph{Racah problem} for $\mathfrak{su}(2)$. The Racah algebra can be def\/ined as the unital associative algebra over $\mathbb{C}$ with generators $\kappa_1$, $\kappa_2$, $\kappa_3$ and relations
\begin{gather}
  [\kappa_1,\kappa_2]=\kappa_3,\qquad
  [\kappa_2,\kappa_3]=a_2\kappa_2^2+a_1\{\kappa_1,\kappa_2\}+c_1 \kappa_1+d \kappa_2 +e_1,\nonumber
  \\
  [\kappa_3,\kappa_1]=a_1\kappa_1^2+a_2\{\kappa_1,\kappa_2\}+c_2 \kappa_2+d \kappa_1+e_2,\label{Racah-1}
 \end{gather}
where $[a,b]=ab-ba$ and $\{a,b\}=ab+ba$ and where $a_i$, $c_i$, $d$, $e_i$ for $i=1,2$ are structure constants. The algebra \eqref{Racah-1} has a Casimir operator
\begin{gather}
 T=a_1\big\{\kappa_1^2,\kappa_2\big\}+a_2\big\{\kappa_1,\kappa_2^2\big\}+\big(a_1^2+c_1\big)\kappa_1^2+\big(a_2^2+c_2\big)\kappa_2^2+\kappa_3^2
\nonumber \\
\hphantom{T=}{}
 +(d+a_1a_2) \{\kappa_1,\kappa_2\}+(2e_1+da_1) \kappa_1+(2e_2+da_2) \kappa_2,\label{Racah-Casimir}
\end{gather}
which commutes with all generators. In~\eqref{Racah-1} only two of the generators are genuinely independent and hence~$\kappa_3$ can be eliminated from the def\/ining relations. Moreover, if $a_1a_2\neq 0$, one can set $a_1=a_2=1$ and $c_1=c_2=0$ by performing af\/f\/ine transformations on the generators~\cite{2014_Genest&Vinet&Zhedanov_LettMathPhys_104_931}.

The Racah algebra can be described as the algebraic structure encoding the bispectrality of the Racah and Wilson polynomials, which sit at the top the Askey scheme of hypergeometric orthogonal polynomials~\cite{2014_Genest&Vinet&Zhedanov_LettMathPhys_104_931}. This assertion originates from the fact that the Racah algebra~\eqref{Racah-1} has realizations on the space of polynomials, called \emph{standard realizations}, in which the generators are the dif\/ference and recurrence operators associated to these polynomials. As an example, in the case of the Racah polynomials $R_n(\lambda(x); \alpha,\beta,\gamma;N)$ (in the notation of~\cite{2010_Koekoek_&Lesky&Swarttouw})  of degree $n$ in the variable $\lambda(x)=x(x+\gamma+\delta+1)$, the standard realization is def\/ined as
\begin{gather}
\label{Racah-SR}
 \kappa_1(x)=B(x) T^{+}+D(x) T^{-}-(B(x)+D(x)) \mathbb{I},\qquad \kappa_2(x)=\lambda(x),
\end{gather}
where $T^{\pm}f(x)=f(x\pm 1)$ are discrete shift operators and where $B(x)$ and $D(x)$ are of the form
\begin{gather*}
 B(x) =\frac{(x+\alpha+1)(x+\beta+\delta+1)(x+\gamma+1)(x+\gamma+\delta+1)}{(2x+\gamma+\delta+1)(2x+\gamma+\delta+2)},
 \\
 D(x) =\frac{x(x-\alpha+\gamma+\delta)(x-\beta+\gamma)(x+\delta)}{(2x+\gamma+\delta)(2x+\gamma+\delta+1)}.
\end{gather*}
The operators \eqref{Racah-SR} satisfy the relations~\eqref{Racah-1} with the structure constants expressed in terms of the parameters~$\alpha$, $\beta$, $\gamma$ and~$N$ \cite{2014_Genest&Vinet&Zhedanov_LettMathPhys_104_931}. In the realization~\eqref{Racah-SR}, the Casimir operator~\eqref{Racah-Casimir} is a~multiple of the identity. In the basis $\{v_i=R_{i}(\lambda(x);\alpha,\beta,\gamma;N)\}_{i=0}^{N}$, the realization~\eqref{Racah-SR} yields a 3-parameter family of $(N+1)$-dimensional irreducible representations of~\eqref{Racah-1} in which $\kappa_1$ is represented by a~diagonal matrix and~$\kappa_2$ is represented by a~tridiagonal matrix. Inf\/inite-dimensional representations of~\eqref{Racah-1} are obtained by taking the standard representation associated with the dif\/ference and recurrence operators of the Wilson polynomials.

The Racah algebra has a an alternative presentation in terms of generators $A$, $B$, $C$ and $\Delta$. In this presentation, one has
\begin{subequations}
 \label{Equitable}
\begin{gather}
\label{First}
 A+B+C=\Gamma, \qquad [A,B]=[B,C]=[C,A]=2\Delta,
\end{gather}
where $\Gamma$ is a constant or a central operator. The commutation relations are
\begin{gather}
 [A,\Delta]=BA-AC+i_{A},\qquad [B,\Delta]=CB-BA+i_{B},\qquad [C,\Delta] =AC-CB+i_{C},
\end{gather}
\end{subequations}
where $i_{A}$, $i_{B}$, $i_{C}$ are structure constants. The details of the passage from~\eqref{Racah-1} to~\eqref{Equitable} can be found in~\cite{2014_Genest&Vinet&Zhedanov_JPhysA_47_025203}. The presentation~\eqref{Equitable} is known as the democratic or equitable presentation, as it is invariant under cyclic permutations of the generators $A$, $B$ and $C$. As far as we know, the democratic presentation of the Racah algebra f\/irst appeared in the work of L\'evy-Leblond and L\'evy-Nahas (equations~(2.8) and (2.9) of~\cite{1965_LevyLeblond&LevyNahas_JMathPhys_6_1372}), who were concerned with the symmetrical coupling of three angular momenta. The algebra~\eqref{Equitable}, which has also been studied in~\cite{2013_Gao&Wang&Hou_LinAlgAppl_439_1834}, can be seen as a~$q=1$ analog of the Askey--Wilson algebra in its $\mathbb{Z}_3$-symmetric presentation, see~\cite{2011_Terwilliger_SIGMA_7_99}.

The Bannai--Ito algebra was f\/irst introduced in 2012 by Tsujimoto, Vinet and Zhedanov in their study of the Bannai--Ito polynomials \cite{2012_Tsujimoto&Vinet&Zhedanov_AdvMath_229_2123}. This algebra can be def\/ined as the unital associative algebra over~$\mathbb{C}$ with generators~$X$, $Y$, $Z$ and relations
\begin{gather}
\label{BI}
 \{X,Y\}=Z+\omega_{Z},\qquad \{Y,Z\}=X+\omega_{X},\qquad \{Z,X\}=Y+\omega_{Y},
\end{gather}
where  $\omega_{X}$, $\omega_{Y}$, $\omega_{Z}$ are structure constants. The Bannai--Ito algebra has a Casimir operator
\begin{gather}
\label{BI-Casimir}
 U=X^2+Y^2+Z^2,
\end{gather}
which commutes with all generators. Again, only two of the generators are genuinely independent. The Bannai--Ito algebra encodes the bispectral properties of the Bannai--Ito polynomials. The standard realization of the Bannai--Ito algebra on the space of polynomials is def\/ined as
\begin{gather}
\label{BI-SR}
 X(z)=F(z) T^{+}R+ G(z) R-(F(z)+G(z)-h)\mathbb{I},\qquad Y(z)=2z+1/2,
\end{gather}
where $Rf(z)=f(-z)$ is the ref\/lection operator, where
\begin{gather*}
  F(z)=\frac{(z-r_1+1/2)(z-r_2+1/2)}{z+1/2},\qquad G(z)=\frac{(z-\rho_1)(z-\rho_2)}{-z},
 \\
 h=\rho_1+\rho_2-r_1-r_2+1/2,
\end{gather*}
and where $\rho_1$, $\rho_2$, $r_1$, $r_2$ are complex parameters. In the realization~\eqref{BI-SR}, the Casimir operator~\eqref{BI-Casimir} acts as a multiple of the identity. If one introduces the basis vectors $w_i=B_{i}(z;\rho_1,\rho_2,r_1,r_2)$ for $i=0,1,2,\ldots$, where $B_{n}(z;\rho_1,\rho_2,r_1,r_2)$ are the Bannai--Ito polynomials of degree $n$ in $z$ (see~\cite{2013_Genest&Vinet&Zhedanov_SIGMA_9_18}), one obtains a four-parameter family of inf\/inite-dimensional irreducible representations of~\eqref{BI} where $X$ is represented by a diagonal matrix and~$Y$ is represented by a tridiagonal matrix; these representations can be made f\/inite-dimensional under appropriate truncation conditions.

The Racah and Bannai--Ito algebras are intimately related to second-order superintegrable systems in two dimensions. Recall that a quantum system in $d$ dimensions governed by a~Hamiltonian~$H$ is said to be superintegrable if it possesses $2d-1$ algebraically independent constants of motion, see~\cite{2013_Miller&Post&Winternitz_JPhysA_46_423001} for a review. On the one hand, the Racah algebra arises as the symmetry algebra for the generic superintegrable system on the two-sphere~\cite{1996_Kalnins&Miller&Pogosyan_JMathPhys_37_6439, 1997_Kalnins&Miller&Pogosyan_JMathPhys_38_5416}. This system is described by the Hamiltonian
\begin{gather}
\label{SI}
 H=L_1^2+L_2^2+L_3^2+\big(x_1^2+x_2^2+x_3^2\big)\left(\frac{k_1^2-1/4}{x_1^2}+\frac{k_2^2-1/4}{x_2^2}+\frac{k_3^2-1/4}{x_3^2}\right),
\end{gather}
where $k_i>-1/2$, $i=1,2,3$, are real parameters and where the operators~$L_i$ are the standard angular momentum generators in three dimensions
\begin{gather*}
 L_1=\frac{1}{i}(x_2 \partial_{x_3}-x_3 \partial_{x_2}),\qquad L_2=\frac{1}{i}(x_3 \partial_{x_1}-x_1 \partial_{x_3}),\qquad L_3=\frac{1}{i}(x_1 \partial_{x_2}-x_2 \partial_{x_1}).
\end{gather*}
The model def\/ined by \eqref{SI} is the most general second-order superintegrable system in two dimensions (without ref\/lections-see below), as all the other systems in that category can be obtained from it by taking limits and contractions~\cite{2013_Kalnins&Miller&Post_SIGMA_9_57}. The connection between the representations of the symmetry algebra generated by the second-order constants of motion of~\eqref{SI} and the Racah/Wilson polynomials was established in~\cite{2007_Kalnins&Miller&Post_JPhysA_40_11525}. On the other hand, the Bannai--Ito algebra arises as the invariance algebra for the system governed by the Hamiltonian~\cite{2014_Genest&Vinet&Zhedanov_JPhysA_47_205202}
\begin{gather}
 \label{SI-2}
 H=L_1^2+L_2^2+L_3^2+\big(x_1^2+x_2^2+x_3^2\big)
 \left(\frac{\mu_1(\mu_1-R_1)}{x_1^2}+\frac{\mu_2(\mu_2-R_2)}{x_2^2}+\frac{\mu_3(\mu_3-R_3)}{x_3^2}\right),
\end{gather}
where $R_if(x_i)=f(-x_i)$, $i=1,2,3$, are ref\/lection operators. Since the Hamiltonian~\eqref{SI-2} commutes with the ref\/lection operators, it can be considered as a concatenation of eight systems of type~\eqref{SI}, which is reminiscent of supersymmetry. The model described by~\eqref{SI-2} is in fact supersymmetric, as it can be expressed, up to additive constant, as the square of a Hermitian supercharge (see~\cite{2014_Genest&Vinet&Zhedanov_JPhysA_47_205202,2015_Genest&Vinet&Zhedanov_CommMathPhys_336_243}).

The fact that the Racah and Bannai--Ito algebras appear as symmetry algebras for the mo\-dels~\eqref{SI} and~\eqref{SI-2} is directly related to the fact that these algebras also arise as the commutant algebras for the actions of the Lie algebra $\mathfrak{su}(1,1)$ and Lie superalgebra $\mathfrak{osp}(1|2)$ on three-fold tensor product representations (see~\cite{2014_Genest&Vinet&Zhedanov_LettMathPhys_104_931} and~\cite{2014_Genest&Vinet&Zhedanov_ProcAmMathSoc_142_1545}). Indeed, in three-fold tensor product representations involving three irreducible representations of the positive-discrete series for $\mathfrak{su}(1,1)$ and $\mathfrak{osp}(1|2)$, the Hamiltonians~\eqref{SI},~\eqref{SI-2} and their symmetries respectively correspond to total and intermediate Casimir operators constructed from the coproducts of~$\mathfrak{su}(1,1)$ and~$\mathfrak{osp}(1|2)$ (see Section~\ref{section3} for details).

In light of the visible relationship between the two Hamiltonians~\eqref{SI},~\eqref{SI-2} and the embedding $\mathfrak{su}(1,1)\subset \mathfrak{osp}(1|2)$, it is natural to expect a relation between the Racah and Bannai--Ito algebras. In the present paper, this relation will be made explicit.

Here is the outline of the paper. In Section~\ref{section2}, we show that upon taking appropriate quadratic combinations of the generators of the Bannai--Ito algebra \eqref{BI}, one f\/inds a central extension of the Racah algebra in its equitable presentation. In Section~\ref{section3}, the quadratic relation between the sCasimir operator of $\mathfrak{osp}(1|2)$ and the Casimir operator of $\mathfrak{su}(1,1)$ is derived. The realization of the Racah generators in terms of the Bannai--Ito generators in the context of three-fold tensor product representations of $\mathfrak{su}(1,1)$ and $\mathfrak{osp}(1|2)$ is given. The relation between the constants of motion of the  superintegrable systems \eqref{SI} and \eqref{SI-2} is deduced. A short conclusion follows.

\section{Embedding of the Racah algebra into the Bannai--Ito algebra\\ and the standard realization}\label{section2}
Consider the standard realization \eqref{BI-SR} of the Bannai--Ito algebra. In this representation, the structure constants have the expressions
\begin{gather*}
 \omega_{X}=4(\rho_1\rho_2+r_1 r_2),\qquad \omega_{Y}=2\big(\rho_1^2+\rho_2^2-r_1^2-r_2^2\big),\qquad \omega_{Z}=4(\rho_1\rho_2-r_1r_2),
\end{gather*}
and the Casimir operator \eqref{BI-Casimir} acts a multiple of the identity with value
\begin{gather*}
 U(z)=u \mathbb{I},\qquad u=2\big(\rho_1^2+\rho_2^2+r_1^2+r_2^2\big)-1/4.
\end{gather*}
Consider the quadratic combination of generators
\begin{gather}
 A(z)=\frac{1}{4}\left(X^2(z)-X(z)-\frac{3}{4}\right),\qquad B(z)=\frac{1}{4}\left(Y^2(z)-Y(z)-\frac{3}{4}\right),
 \nonumber\\
 C(z)=\frac{1}{4}\left(Z^2(z)-Z(z)-\frac{3}{4}\right),\label{Combinations}
 \end{gather}
and introduce the operator
\begin{gather}
\label{A1}
 I(z)=X(z)+Y(z)+Z(z)-3/2.
\end{gather}
It is verif\/ied that $I(z)$ commutes the operators $A(z)$, $B(z)$ and $C(z)$, i.e.,
\begin{gather*}
 [A(z), I(z)]=[B(z), I(z)]=[C(z), I(z)]=0.
\end{gather*}
It is also observed that the generators $A(z)$, $B(z)$, $C(z)$ commute with their sum since one has
\begin{gather}
\label{A2}
 A(z)+B(z)+C(z)=\frac{1}{4} [U(z)-I(z)-15/4 ],
\end{gather}
where $U(z)$ is the Bannai--Ito Casimir operator~\eqref{BI-Casimir}. Let $\Delta(z)$ be def\/ined as
\begin{gather*}
 [A(z),B(z)]=[B(z), C(z)]=[C(z), A(z)]=2\Delta(z).
\end{gather*}
A direct calculation using the expressions \eqref{BI-SR} for the standard realization shows that the operators $A(z)$, $B(z)$, $C(z)$ and $\Delta(z)$ satisfy the commutation relations
\begin{gather}
 [A(z), \Delta(z)] =B(z) A(z)-A(z) C(z)+\frac{1}{16}\left(\frac{\omega_{Y}-\omega_{Z}}{2}\right)\left[\left(\frac{\omega_{Y}+\omega_{Z}}{2}\right)-I(z)\right],
 \nonumber\\
 [B(z), \Delta(z)] =C(z) B(z)-B(z) A(z)+\frac{1}{16}\left(\frac{\omega_{Z}-\omega_{X}}{2}\right)\left[\left(\frac{\omega_{Z}+\omega_{X}}{2}\right)-I(z)\right],
 \label{Relations}\\
  [C(z), \Delta(z)] =A(z) C(z)-C(z) B(z)+\frac{1}{16}\left(\frac{\omega_{X}-\omega_{Y}}{2}\right)\left[\left(\frac{\omega_{X}+\omega_{Y}}{2}\right)-I(z)\right].\nonumber
  \end{gather}
Upon comparing \looseness=-1 the above relations with those appearing in~\eqref{Equitable}, it is directly seen that the above relations coincide with the def\/ining relations of the equitable Racah algebra extended by the central operator $I(z)$. We have thus shown that the quadratic combinations \eqref{Combinations} of the generators of the Bannai--Ito algebra~\eqref{BI} in the standard realization~\eqref{BI-SR} realize the Racah algebra. The explicit expression of the generators $A(z)$, $B(z)$ and $C(z)$ should be compared with~\eqref{Racah-SR}.

\begin{remark}
 While the above calculations have been performed with the help of the faithful model provided by the realization~\eqref{BI-SR}, let us observe that the embedding \eqref{Relations} of the Racah algebra into the Bannai--Ito algebra can be obtained directly in an abstract fashion using only the def\/initions~\eqref{Combinations},~\eqref{A1}, \eqref{A2}, and the relations~\eqref{BI}.
\end{remark}

\section{Embedding of the Racah algebra into the Bannai--Ito algebra,\\ superintegrability in two dimensions and the Racah problem}\label{section3}

In this section we give another embedding of the Racah algebra into the Bannai--Ito algebra in relation with the realizations of these algebras as symmetry algebras for the superintegrable systems~\eqref{SI} and~\eqref{SI-2} or equivalently as commutant algebras for the actions of $\mathfrak{su}(1,1)$ and $\mathfrak{osp}(1|2)$ on three-fold tensor product representations. Let us f\/irst review how the Racah and Bannai--Ito algebras arise in the context of the latter.

\subsection[The Racah algebra and the Lie algebra $\mathfrak{su}(1,1)$]{The Racah algebra and the Lie algebra $\boldsymbol{\mathfrak{su}(1,1)}$}

The $\mathfrak{su}(1,1)$ algebra has three generators $K_{0}$, $K_{\pm}$ satisfying the commutation relations
\begin{gather*}
 [K_0,K_{\pm}]=\pm K_{\pm},\qquad [K_{-},K_{+}]=2K_0.
\end{gather*}
The Casimir operator for $\mathfrak{su}(1,1)$ can be written as
\begin{gather}
\label{SU-Cas}
 C=K_0^2-K_{+}K_{-}-K_0,
\end{gather}
As a Lie algebra, $\mathfrak{su}(1,1)$ admits a coproduct $\Delta\colon \mathfrak{su}(1,1)\rightarrow \mathfrak{su}(1,1)\otimes \mathfrak{su}(1,1)$ given by
\begin{gather}
\label{SU-Delta}
 \Delta(K_{\pm})=1\otimes K_{\pm}+K_{\pm}\otimes 1,\qquad \Delta(K_0)=1\otimes K_0+K_0\otimes 1.
\end{gather}
The coproduct allows to construct tensor product representations. Let $V^{(\lambda)}$ be an irreducible representation space for $\mathfrak{su}(1,1)$ on which the Casimir operator $C$ acts as a $\lambda \mathbb{I}$. Consider the three-fold tensor product representation on $V=V^{(\lambda_1)}\otimes V^{(\lambda_2)}\otimes V^{(\lambda_3)}$.  The action of the generators is given by
\begin{gather}
\label{Action-2}
 K_{\pm}V=(1\otimes \Delta)\Delta(K_{\pm}) V,\qquad K_{0}V=(1\otimes \Delta)\Delta(K_{0}) V.
\end{gather}
The elements that commute with the action of $\mathfrak{su}(1,1)$ on $V$ are easily constructed. First, one has the three initial Casimir operators $C^{(i)}$, $C^{(2)}$, $C^{(3)}$ for $i=1,2,3$ associated to each factor~$V^{(\lambda_i)}$ of the tensor product. These operators are def\/ined as
\begin{gather*}
 C^{(1)}=C\otimes 1\otimes 1,\qquad  C^{(2)}=1\otimes C\otimes 1,\qquad  C^{(3)}=1\otimes 1\otimes C.
\end{gather*}
Second, one has the total Casimir operator $C^{(4)}$ that has the expression
\begin{gather*}
C^{(4)}=(1\otimes \Delta)\Delta(C)=(\Delta\otimes 1)\Delta(C),
\end{gather*}
and which acts non-trivially on all factors of the tensor product. Finally, one has the two intermediate Casimir operators~$C^{(12)}$ and~$C^{(23)}$ that are of the form
\begin{gather*}
 C^{(12)}=\Delta(C)\otimes 1\qquad C^{(23)}=1\otimes \Delta(C),
\end{gather*}
and which act non-trivially only on a pair of representation spaces. By construction, the initial, intermediate and total Casimir operators commute with action~\eqref{Action-2} of the~$\mathfrak{su}(1,1)$ generators on the tensor product space $V$. However, these Casimir operators do not all commute with each other and they thus generate a non commutative commutant algebra. This algebra can be identif\/ied with the Racah algebra as follows.

Let $\mathcal{A}$ and $\mathcal{C}$ be def\/ined as
$\mathcal{A}=C^{(23)}$, $\mathcal{C}=C^{(12)}$,
and let $ \mathcal{B}=\sum\limits_{i=1}^{4}C^{(i)}-\mathcal{A}-\mathcal{C}$. By def\/inition, we thus have
\begin{gather*}
 \mathcal{A}+\mathcal{B}+\mathcal{C}=C^{(1)}+C^{(2)}+C^{(3)}+C^{(4)}.
\end{gather*}
Since $\mathcal{A}$, $\mathcal{B}$ and $\mathcal{C}$ all commute with the Casimir operators $C^{(i)}$ for $i=1,2,3,4$, one can introduce the operator $\Omega$ def\/ined as
\begin{gather*}
 [\mathcal{A},\mathcal{B}]=[\mathcal{B},\mathcal{C}]=[\mathcal{C},\mathcal{A}]\equiv 2 \Omega.
\end{gather*}
Using the expressions of the intermediate and total Casimir operators computed from \eqref{SU-Cas} and the coproduct~\eqref{SU-Delta}, an involved calculation shows that the operators~$\mathcal{A}$, $\mathcal{B}$, $\mathcal{C}$ together with the operator~$\Omega$, satisfy the commutation relations
\begin{gather}
  [\mathcal{A},\Omega] =\mathcal{B}\mathcal{A}-\mathcal{A}\mathcal{C}+(\lambda_2-\lambda_3)(\lambda_4-\lambda_1),
\qquad
   [\mathcal{B},\Omega] =\mathcal{C}\mathcal{B}-\mathcal{B}\mathcal{A}+(\lambda_3-\lambda_1)(\lambda_4-\lambda_2),
 \nonumber  \\
   [\mathcal{C},\Omega] =\mathcal{A}\mathcal{C}-\mathcal{C}\mathcal{B}+(\lambda_1-\lambda_2)(\lambda_4-\lambda_3),\label{Racah-2}
 \end{gather}
where the substitutions $C^{(i)}=\lambda_i$ for $i=1,2,3,4$ were made to underscore the fact that each eigenspace of $C^{(4)}$ supports a representation of the Racah algebra~\eqref{Racah-2}.

The relationship between the results presented above and the analysis of the generic superintegrable system on the 2-sphere is as follows~\cite{2014_Genest&Vinet&Zhedanov_LettMathPhys_104_931}. Let $V^{(k)}$ be the representation associated with the following realization of $\mathfrak{su}(1,1)$:
\begin{gather}
\label{rea}
 K_0(x)=\frac{1}{4}\left(-\partial_{x}^2+x^2+\frac{k^2-1/4}{x^2}\right),\qquad K_{\pm}(x)=\frac{1}{4}\left((x\mp \partial_{x})^2-\frac{k^2-1/4}{x^2}\right),
\end{gather}
where $k$ is a real number such that $k>-1/2$. In this realization, one has
\begin{gather*}
 C(x)=\frac{1}{4}\big(k^2-1\big) \mathbb{I}.
\end{gather*}
This realization corresponds to positive-discrete series \cite{1991_Vilenkin&Klimyk}. Consider the tensor product representation $V^{(k_1)}\otimes V^{(k_2)}\otimes V^{(k_3)}$ obtained by taking three copies of the realization \eqref{rea} with parameters $k_1$, $k_2$ and $k_3$ in the three independent variables $x_1$, $x_2$ and $x_3$.  In this realization, the total Casimir operator $C^{(4)}$ takes the form
\begin{gather}
 C^{(4)}(x_1,x_2,x_3)=\frac{1}{4}\Bigg(L_1^2+L_2^2+L_3^2
 \nonumber\\
 \hphantom{C^{(4)}(x_1,x_2,x_3)=\frac{1}{4}\Bigg(}{}
 +\big(x_1^2+x_2^2+x_3^2\big) \left(\frac{k_1^2-1/4}{x_1^2}+\frac{k_2^2-1/4}{x_2^2}+\frac{k_3^2-1/4}{x_3^2}\right)-\frac{3}{4}\Bigg).\label{Ultra}
\end{gather}
Upon comparing with \eqref{SI}, it is directly seen that the total Casimir operator corresponds, up to an inessential af\/f\/ine transformation, to the Hamiltonian for the generic superintegrable system on the two-sphere. By construction, the intermediate Casimir operators~$\mathcal{A}$, $\mathcal{B}$ and~$\mathcal{C}$ commute with $C^{(4)}$, and are thus the constants of motion of the Hamiltonian. In this realization, one has
\begin{gather}
 \mathcal{A}(x_1,x_2,x_3) =\frac{1}{4}\left[L_1^2+\big(x_2^2+x_3^2\big)\left(\frac{k_2^2-1/4}{x_2^2}+\frac{k_3^2-1/4}{x_3^2}\right)-1\right],\nonumber\\
  \mathcal{B}(x_1,x_2,x_3) =\frac{1}{4}\left[L_2^2+\big(x_1^2+x_3^2\big)\left(\frac{k_3^2-1/4}{x_3^2}+\frac{k_1^2-1/4}{x_1^2}\right)-1\right],\label{Realization-1}\\
   \mathcal{C}(x_1,x_2,x_3) =\frac{1}{4}\left[L_3^2+\big(x_1^2+x_2^2\big)\left(\frac{k_1^2-1/4}{x_1^2}+\frac{k_2^2-1/4}{x_2^2}\right)-1\right].\nonumber
   \end{gather}
The constants of motion $\mathcal{A}(x_1,x_2,x_3)$, $\mathcal{B}(x_1,x_2,x_3)$ and $\mathcal{C}(x_1,x_2,x_3)$ satisfy the def\/ining relations of the equitable Racah algebra \eqref{Racah-2} with $\lambda_i=(k_i^2-1/4)/4$ for $i=1,2,3$ and $\lambda_4=(H-3/4)/4$ where $H$ is the Hamiltonian \eqref{SI} of the generic superintegrable system on the 2-sphere. The constants of motion~\eqref{Realization-1} coincide with those constructed by Miller et al.~\cite{1996_Kalnins&Miller&Pogosyan_JMathPhys_37_6439, 1997_Kalnins&Miller&Pogosyan_JMathPhys_38_5416,2013_Kalnins&Miller&Post_SIGMA_9_57}.

\subsection[The Bannai--Ito algebra and the Lie superalgebra $\mathfrak{osp}(1|2)$]{The Bannai--Ito algebra and the Lie superalgebra $\boldsymbol{\mathfrak{osp}(1|2)}$}

The $\mathfrak{osp}(1|2)$ Lie superalgebra has two odd generators $A_{\pm}$ and one even generator $A_0$ that obey the commutation relations
\begin{gather*}
 [A_0, A_{\pm}]=\pm \frac{A_{\pm}}{2},\qquad \{A_{+},A_{-}\}=2A_0.
\end{gather*}
The even subalgebra of $\mathfrak{osp}(1|2)$  is generated by the elements~$J_{\pm}=A_{\pm}^2$ and~$A_0$. The generators of the even subalgebra satisfy the def\/ining relations of~$\mathfrak{su}(1,1)$. One has indeed
\begin{gather*}
 [A_0,J_{\pm}]=\pm J_{\pm},\qquad [J_{-},J_{+}]=2A_0,
\end{gather*}
whence it follows that $\mathfrak{su}(1,1)\subset \mathfrak{osp}(1|2)$. The abstract $\mathbb{Z}_2$ grading of $\mathfrak{osp}(1|2)$ can be concretized by appending the grade involution $P$ to the set of generators and declaring that the even generators commute with $P$ and that the odd generators anticommute with $P$. Hence one can def\/ine $\mathfrak{osp}(1|2)$ as the algebra generated by $A_{\pm}$, $A_0$ and the involution $P$ satisfying the relations
\begin{gather*}
 [A_0, A_{\pm}]=\pm \frac{A_{\pm}}{2},\qquad \{A_{+},A_{-}\}=2A_0,\qquad [A_0, P]=0\qquad \{A_{\pm},P\}=0,\qquad P^2=1.
\end{gather*}
This presentation has been referred to as $sl_{-1}(2)$ \cite{2011_Tsujimoto&Vinet&Zhedanov_SIGMA_7_93}; in this presentation the grading of the elements no longer needs to be specif\/ied. The sCasimir operator~$S$ of $\mathfrak{osp}(1|2)$ is def\/ined as~\cite{1995_Lesniewski_JMathPhys_36_1457}
\begin{gather*}
 S=2A_{+}A_{-}-2A_0+1/2.
\end{gather*}
The sCasimir operator satisf\/ies the relations
\begin{gather*}
 \{S,A_{\pm}\}=0,\qquad [S,A_{0}]=0.
\end{gather*}
There is a quadratic relation between the sCasimir of~$\mathfrak{osp}(1|2)$ and the Casimir operator of the even~$\mathfrak{su}(1,1)$ subalgebra, denoted by $C_{\mathfrak{su}(1,1)\subset \mathfrak{osp}(1|2)}$. A direct calculation shows that
\begin{gather}
\label{Crucial}
 C_{\mathfrak{su}(1,1)\subset \mathfrak{osp}(1|2)}=\frac{1}{4}\left(S^2+S-\frac{3}{4}\right),
\end{gather}
where one has
\begin{gather*}
 C_{\mathfrak{su}(1,1)\subset \mathfrak{osp}(1|2)}=A_{0}^2-J_{+}J_{-}-A_0,
\end{gather*}
as per \eqref{SU-Cas}. An operator that commutes with every element of~$\mathfrak{osp}(1|2)$ can be constructed by combining the sCasimir with the grade involution. It is indeed verif\/ied that the operator
\begin{gather}
\label{Cas-OSP}
 Q=S\cdot P=2A_{+}A_{-}P-2A_0P+P/2,
\end{gather}
commutes with $A_{\pm}$ and $A_0$. We shall refer to \eqref{Cas-OSP} as the Casimir operator for~$\mathfrak{osp}(1|2)$. The~$\mathfrak{osp}(1|2)$ algebra has a coproduct def\/ined as follows
\begin{gather}
\label{Delta-OSP}
 \Delta(A_{\pm})=A_{\pm}\otimes P+1\otimes A_{\pm},\qquad \Delta(A_0)=1\otimes A_0+A_0\otimes 1,\qquad \Delta(P)=P\otimes P.
\end{gather}
This coproduct allows to construct tensor product representations. Let $W^{(\lambda)}$ be an irreducible representation space for $\mathfrak{osp}(1|2)$ on which the Casimir operator \eqref{Cas-OSP} acts as~$\lambda \mathbb{I}$. Consider the three-fold tensor product representation $W=W^{(\lambda_1)}\otimes W^{(\lambda_2)}\otimes W^{(\lambda_3)}$. The action of the~$\mathfrak{osp}(1|2)$ generators is \begin{gather*}
 A_{\pm}W=(1\otimes \Delta)\Delta(A_{\pm}) W,\qquad A_{0}W=(1\otimes \Delta)\Delta(A_{0}) W,\qquad P W=(1\otimes \Delta)\Delta(P) W.
\end{gather*}
Once again, the elements that commute with this action are easily found. First, one has the initial Casimir operators $Q^{(i)}$, $Q^{(2)}$, $Q^{(3)}$ for $i=1,2,3$ associated to each factor $W^{(\lambda_i)}$. They are def\/ined as
\begin{gather*}
 Q^{(1)}=Q\otimes 1\otimes 1,\qquad  Q^{(2)}=1\otimes Q\otimes 1,\qquad  Q^{(3)}=1\otimes 1\otimes Q.
\end{gather*}
Second, one has the total Casimir operator $Q^{(4)}$ def\/ined as
\begin{gather*}
Q^{(4)}=(1\otimes \Delta)\Delta(Q)=(\Delta\otimes 1)\Delta(Q),
\end{gather*}
which acts non-trivially on all factors of the three-fold tensor product. Finally, one has the two intermediate Casimir operators~$Q^{(12)}$ and $Q^{(23)}$ given by
\begin{gather*}
 Q^{(12)}=\Delta(Q)\otimes 1\qquad Q^{(23)}=1\otimes \Delta(Q),
\end{gather*}
which act non-trivially only on a pair of representation spaces. By construction, the initial, intermediate and total Casimir operators commute with the action of the $\mathfrak{osp}(1|2)$ generators on the tensor product space $W$. However, these Casimir operators do not all commute with each other and they thus give rise to a non-commutative commutant algebra. This algebra can be identif\/ied with the Bannai--Ito algebra in the following way.

Let $Z$ and $X$ be def\/ined by
\begin{gather*}
 Z=-Q^{(12)},\qquad X=-Q^{(23)}.
\end{gather*}
Upon using the formulas~\eqref{Cas-OSP} and~\eqref{Delta-OSP} for the Casimir and coproduct for~$\mathfrak{osp}(1|2)$, an involved calculation shows that the operators~$Z$, $X$ satisfy the relations
\begin{gather}
\label{Comu}
 \{Z, X\}=Y+\omega_{Y},\qquad \{X,Y\}=Z+\omega_{Z},\qquad \{Y,Z\}=X+\omega_{X},
\end{gather}
where $Y$ is def\/ined by the f\/irst relation of \eqref{Comu} and where
\begin{gather}
\label{Struc}
 \omega_{X}=2(\lambda_2\lambda_3+\lambda_1\lambda_4),\qquad \omega_{Y}=2(\lambda_1\lambda_3+\lambda_2\lambda_4),\qquad \omega_{Z}=2(\lambda_1\lambda_2+\lambda_3\lambda_4).
\end{gather}
Again we have made the substitutions $Q^{(i)}\rightarrow -\lambda_i$ for $i=1,2,3,4$, replacing the initial and total Casimir operators, which commute with $X$, $Y$, $Z$, by constants.

The relationship between the superintegrable system with ref\/lections on the two-sphere governed by the Hamiltonian~\eqref{SI-2} and the above considerations can be established as follows. Let~$W^{(\mu)}$ be the representation of $\mathfrak{osp}(1|2)$  associated with the coordinate realization
\begin{gather}
 A_{\pm}(x)=\frac{1}{2}\left(x\mp \partial_{x}\pm \frac{\mu}{x} R_{x}\right),\qquad A_0(x)=\frac{1}{4}\left(-\partial_{x}^2+x^2+\frac{\mu (\mu-R_{x})}{x^2}\right),\nonumber\\
  P(x)=R_{x},\label{dompe}
\end{gather}
where $R_x f(x)=f(-x)$ is the ref\/lection operator and where $\mu$ is a real parameter such that~$\mu>0$. In this realization, the Casimir operator $Q$ of $\mathfrak{osp}(1|2)$ acts as a multiple of the identity with multiple
\begin{gather*}
 Q(x)=-\mu \mathbb{I}.
\end{gather*}
The Casimir operator for the even $\mathfrak{su}(1,1)$ subalgebra has the expression
\begin{gather}
\label{exp}
 C_{\mathfrak{su}(1,1)\subset \mathfrak{osp}(1|2)}=A_0(x)^2-J_{+}(x)J_{-}(x)-A_0(x)=\frac{1}{4}\left[\mu^2-\mu R_{x}-\frac{3}{4}\right].
\end{gather}
Consider now the tensor product representation $W=W^{(\mu_1)}\otimes W^{(\mu_2)}\otimes W^{(\mu_3)}$ obtained by combining three copies of the realization~\eqref{dompe} in three independent variables~$x_1$,~$x_2$,~$x_3$ using the coproduct specif\/ied in~\eqref{Delta-OSP}. In this realization one has
\begin{gather}
 \big[S^{(4)}(x_1,x_2,x_3)\big]^{2}+S^{(4)}(x_1,x_2,x_3)\label{H2}
\nonumber \\
\qquad{}= L_1^2+L_2^2+L_3^2+\big(x_1^2+x_2^2+x_3^2\big)\left(\frac{\mu_1(\mu_1-R_1)}{x_1^2}
+\frac{\mu_2(\mu_2-R_2)}{x_2^2}+\frac{\mu_3(\mu_3-R_3)}{x_3^2}\right),\nonumber
\end{gather}
where $S^{(4)}$ is the total sCasimir operator. It is directly seen that the right-hand side of~\eqref{H2} corresponds to the Hamiltonian for the generic superintegrable system on the 2-sphere with ref\/lections. The supersymmetry of~\eqref{H2} is manifest from the above equation. Indeed, the right-hand side of~\eqref{H2} can be written, up to a constant, as the square of the Hermitian super\-char\-ge~$S^{(4)}$. The intermediate Casimir operators~$X$ and~$Z$ as well as the derived operator $Y$ commute with the total Casimir~$Q^{(4)}$, but also with the total sCasimir~$S^{(4)}$. Indeed, in view of~\eqref{Cas-OSP} and~\eqref{Delta-OSP}, one has
\begin{gather*}
 S^{(4)}(x_1,x_2,x_3)=Q^{(4)}(x_1,x_2,x_3)\cdot R_1 R_2 R_3.
\end{gather*}
Since the operators $X$, $Y$, $Z$ commute with the product $R_1R_2R_3$, it follows that they are symmetries of~$S^{(4)}$ and consequently of the Hamiltonian~\eqref{SI-2}. In the present realization, the constants of motion have the expressions
 \begin{gather}
  X(x_1,x_2,x_3) =\left(i L_1+\mu_2 \frac{x_3}{x_2} R_2-\mu_3 \frac{x_2}{x_3}R_3\right)R_2+\mu_2 R_3+\mu_3 R_2+R_2 R_3/2,
 \nonumber \\
  Y(x_1,x_2,x_3) =\left(-i L_2+\mu_1 \frac{x_3}{x_1} R_1-\mu_3 \frac{x_1}{x_3}R_3\right)R_1R_2+\mu_1 R_3+\mu_3 R_1+R_1 R_3/2,
  \label{Realization-2}\\
  Z(x_1,x_2,x_3) =\left(i L_3+\mu_1 \frac{x_2}{x_1} R_1-\mu_2 \frac{x_1}{x_2}R_2\right)R_1+\mu_1 R_2+\mu_2 R_1+R_1 R_2/2.
  \nonumber
 \end{gather}
They satisfy the commutations relations~\eqref{Comu} with the structure constants~\eqref{Struc}. In this realization, the Casimir operator of the Bannai--Ito algebra has the expression
\begin{gather*}
 U(x_1,x_2,x_3)=\big[S^{(4)}\big]^2+\mu_1^2+\mu_2^2+\mu_3^2-1/4.
\end{gather*}

\subsection{Relationship between the two realizations}
Let us now make explicit the relationship between the realization~\eqref{Realization-1} of the Racah algebra~\eqref{Racah-2} by the constants of motion of the generic superintegrable system on the 2-sphere governed by the Hamiltonian~\eqref{Ultra} and the realization~\eqref{Realization-2} of the Bannai--Ito algebra~\eqref{Comu} by the joint symmetries of the total sCasimir operator~$S^{(4)}$ and the generic superintegrable system with ref\/lections on the 2-sphere described by the Hamiltonian~\eqref{H2}. In view of the relation~\eqref{Crucial}, we consider the quadratic combinations
\begin{gather}
 A(x_1,x_2,x_3)=\frac{1}{4}\left[X^2(x_1,x_2,x_3)-X(x_1,x_2,x_3)R_{2}R_{3}-\frac{3}{4}\right],
\nonumber \\
 B(x_1,x_2,x_3)=\frac{1}{4}\left[Y^2(x_1,x_2,x_3)-Y(x_1,x_2,x_3)R_{3}R_{1}-\frac{3}{4}\right],\label{A3}
 \\
 C(x_1,x_2,x_3)=\frac{1}{4}\left[Z^2(x_1,x_2,x_3)-Z(x_1,x_2,x_3)R_{1}R_{2}-\frac{3}{4}\right].
\nonumber
\end{gather}

\begin{remark}
 Let us note that the combinations \eqref{A3} dif\/fer from those considered in \eqref{Combinations} since they involve the ref\/lection operators $R_1$, $R_2$ and $R_3$. This is related to the fact that the opera\-tors~\eqref{Realization-2} realize in this case a central extension of the Bannai--Ito algebra with the total Casimir operator~$Q^{(4)}(x_1,x_2,x_3)$ arising in the structure ``constants'', see~\eqref{Struc}.
\end{remark}

A direct calculation shows that the operators $A$, $B$ and $C$ have the expressions
\begin{gather}
 A(x_1,x_2,x_3) =\frac{1}{4}\left[L_1^2+\big(x_2^2+x_3^2\big)\left(\frac{\mu_2(\mu_2-R_2)}{x_2^2}+\frac{\mu_3(\mu_3-R_3)}{x_3^2}\right)-1\right],
\nonumber \\
  B(x_1,x_2,x_3) =\frac{1}{4}\left[L_2^2+\big(x_3^2+x_1^2\big)\left(\frac{\mu_3(\mu_3-R_3)}{x_3^2}+\frac{\mu_1(\mu_1-R_1)}{x_1^2}\right)-1\right],
\label{vlan}  \\
  C(x_1,x_2,x_3) =\frac{1}{4}\left[L_3^2+\big(x_1^2+x_2^2\big)\left(\frac{\mu_1(\mu_1-R_1)}{x_1^2}+\frac{\mu_2(\mu_2-R_2)}{x_2^2}\right)-1\right].
\nonumber
  \end{gather}
These expressions are seen to correspond to the ones in~\eqref{Realization-1} with $k_i$ having the expression
\begin{gather}
\label{ppp}
 k_i=(\mu_i-R_i/2),\qquad i=1,2,3.
\end{gather}
It is directly verif\/ied that the operators $A(x_1,x_2,x_3)$, $B(x_1,x_2,x_3)$ and $C(x_1,x_2,x_3)$ commute with all the ref\/lections. As a consequence, their algebraic properties are not af\/fected by the operator-valuedness of the~$k_i$. It follows that the operators \eqref{vlan} provide a realization of the Racah algebra~\eqref{Racah-2} where
\begin{gather*}
 \lambda_i=\frac{1}{4}\left(\mu_i^2-\mu_i R_i-\frac{3}{4}\right),
\end{gather*}
for $i=1,2,3$ and where $\lambda_4$ corresponds to the Hamiltonian \eqref{Ultra} with the $k_i$ as in~\eqref{ppp}.

\section{Conclusion}\label{section4}

In this paper we have provided embeddings of the Racah into the Bannai--Ito algebra in two dif\/ferent realizations. On the one hand, we have shown that in the standard realization of the Bannai--Ito algebra on the space of polynomials, certain quadratic combinations of the Bannai--Ito generators realize a central extension of the Racah algebra. On the other hand, we have shown that in the realization of the Bannai--Ito algebra by the constants of motion of the generic second-order superintegrable on the 2-sphere with ref\/lections, again quadratic combinations of the Bannai--Ito generators involving ref\/lections satisfy the Racah algebra relations with structure constants that depend on ref\/lections.

The results presented here establish that the Racah algebra is embedded into the Bannai--Ito algebra. This supports the assertion that the Bannai--Ito algebra can in fact be considered as more fundamental than the Racah algebra, the former being essentially the ``square-root'' of the latter. In view of the deep relationship between the Racah and Bannai--Ito algebras with the Racah/Wilson and Bannai--Ito polynomials, it should be possible, in principle, to derive the main properties of the Racah/Wilson polynomials from those of the Bannai--Ito polynomials. It would be of interest to investigate this possibility.

\subsection*{Acknowledgements}
The authors would like to thank an anonymous referee for pointing out that the embedding of the Racah algebra in the Bannai--Ito algebra proposed in section two holds in the abstract. V.X.G.~is supported by the Natural Science and Engineering Research Council of Canada (NSERC). The research of L.V.~is supported in part by NSERC. A.Z.~wishes to thank the Centre de Recherches Math\'ematiques for its hospitality.

\pdfbookmark[1]{References}{ref}
\LastPageEnding

\end{document}